\documentstyle[prc,preprint,aps,epsfig]{revtex}                                 
\begin{document}                                                                
\title{Theoretical Predictions for Pionium Searches}                            
\author{M.\ Sander, C.\ Kuhrts and H.\ V.\ von Geramb}                          
\address{Theoretische Kernphysik, Universit\"at Hamburg,                        
D--22761 Hamburg, Germany}                                                      
\date{\today}                                                                   
\maketitle                                                                      
\begin{abstract}                                                                
Characteristic properties of pionium $A_{2\pi}$ and associated                  
low energy s--wave                                                              
cross sections $\sigma(\pi^0\pi^0\to \pi^0\pi^0)$,                              
$\sigma(\pi^+\pi^-\to\pi^0\pi^0)$ and $\sigma(\pi^0\pi^0\to \pi^+\pi^-)$        
are investigated with a coupled channels potential model. Some                  
experimental results and conclusions are to be reconsidered.                    
\end{abstract}                                                                  
Like other charged particle pairs the dimesonic $\pi^+\pi^-$ system             
is expected to form a Coulomb bound atomic system                               
known as pionium $A_{2\pi}$.  Pionium is realized by Coulomb attraction         
and their detailed properties                                                   
are  affected by the hadronic  interaction.                                     
It has  a finite lifetime due to                                                
strong coupling into the energetically open $\pi^0\pi^0$                        
channel and pion decays in general.                                             
                                                                                
We are treating                                                                 
pionium  as a low energy s--wave coupled channel resonance                      
with a coupled channels potential model                                         
\begin{equation} \label{eqn1}                                                   
f_i^{\prime\prime} + k_i^2 f_i = \sum_{j=1}^2 m_iV_{ij} f_j,                    
\quad i=1(2)\mbox{\ \ for\ \ }                                                  
\pi^+\pi^-(\pi^0\pi^0).\end{equation}                                           
with $M_{\pi\pi}=2\sqrt{k_i^2+m_i^2}$ and the interaction                       
potential                                                                       
\begin{eqnarray}                                                                
 V_{11}  & = &  V^{\pi^+\pi^-} =                                                
\frac13 V_0^2 + \frac23 V_0^0 - \frac{e^2}{r}, \\                               
V_{12}  = V_{21}  & = & {\sqrt{2}\over 3}(V_0^2 - V_0^0), \\                    
V_{22}  & = & V^{\pi^0\pi^0} = \frac23 V_0^2 + \frac13 V_0^0.                   
\end{eqnarray}                                                                  
The potential matrix is determined with                                         
 results of quantum inversion  of the $T=0,\, 2$                                
experimental and theoretical hadronic phase shifts                              
$\delta_0^T(M_{\pi\pi})$ in the elastic domain                                  
for $M_{\pi\pi}\le 1$ GeV respectively \cite{fro77,loh90,ger94}.                
Symmetriced mesonic wave functions and the isospin projectors                   
\begin{equation}                                                                
                 P(T=0)={1-\tau.\tau\over 3},\quad                              
                 P(T=2)={2+\tau.\tau\over 3},                                   
\end{equation}                                                                  
are used to construct the potential matrix (2--4).                              
              The input phase shift function and                                
Froggatt data \cite{fro77} are shown in Fig. 1 with the implication that        
Lohse data \cite{loh90} give a qualitative similar picture.                  
Gelfand--Levitan--Machenko inversion potentials for                             
Froggatt and Lohse inputs                                                       
as well as the potential matrix  is shown in Fig. 2. We verified that           
the inversion potentials are reproducing for all energies                       
the input phase functions within 0.1 degrees and                                
scattering lengths in Table \ref{tab_pipi}.                                     
With numerical integration of (\ref{eqn1}) and asymptotic matching              
to Coulomb and Bessel functions  we determine                                   
the normalization constants (below threshold for the closed $\pi^+\pi^-$        
channel) and the hadronic S--matrix.                                            
The pure Coulomb potential supports in principle an infinite set                
$ |\pi^+\pi^-,nS>, n=1,2,\dots$ of bound states. We take for the                
$\pi^+\pi^-$ asymptotic wave function                                           
a finite superposition of states. In particular  we study the                   
pure ground state resonance, which we call the {\em pionium--proper}            
with n=1, and excited states with  n=2--5. The                                  
elastic $\pi^0\pi^0$  cross sections with the pionium--proper resonance         
is shown in Fig. 3 and reaction cross sections in Fig. 4.                       
They are determined with standard expressions from the S-matrix                 
\begin{equation}                                                                
\sigma(\pi^{00}\to \pi^{00})={\pi\over k_2^2}|1-S_{22}|^2                       
\end{equation}                                                                  
and                                                                             
\begin{eqnarray}                                                                
\sigma(\pi^{00}\to \pi^{+-}) & = & {\pi\over k_2^2}(1-|S_{22}|^2),\\            
\sigma(\pi^{+-} \to \pi^{00}) & = & {\pi\over k_1^2}(1-|S_{11}|^2).             
\end{eqnarray}                                                                  
The pionium--proper resonance in the $\pi^0\pi^0$ channel,                      
Fig. 3 (top), has a FWHM = 14$^F$(20$^L$) eV                                   
equivalent to a lifetime $\tau = 4.7^F (3.3^L) \times 10^{-17}$ sec.            
Its binding energy is $E_B = 2.445^F (2.407^L)$ keV in contrast to              
$E_B = 1.858$ keV for the 1S Coulomb ground state.                              
The pionium resonance cross sections are large (4 barns)                        
and     equal in magnitude for all resonances                                   
$n=1-5$.                                                                        
Their resonance widths are small and they follow very well the rule             
$\Gamma_1/n^2$ eV, see Table \ref{tab_pipi}.                                    
Adjacent  to the resonances we find in the continuum                            
as prominent and important feature a large threshold                            
reaction cross section                                                          
peak for $\pi^+\pi^-\to \pi^0\pi^0$  which is shown in                          
Fig. 4. Like the pionium resonances  this peak has a large cross section        
($\sim$barn) and a width of $\sim$MeV.                                          
                                                                                
We are aware of the claim that pionium has been seen in  high energy            
experiments \cite{afa93} and  continued interest exists by                      
experimentalists and theoreticians \cite{poc94,kur94}. However,                 
on the basis of our presented numerical results                                 
and the experiment description we claim that                                    
pionium does                                                                    
not necessarily explain the data in experiment \cite{afa93}.  Rather,           
the experimental $\pi^+\pi^-$ surplus is an effect of  $\pi^+,\pi^-$            
{\em initial state interactions} with the strong Coulomb field of target        
nuclei and the strong  kinematically caused energy dependence                   
of the $\pi^+\pi^- \to \pi^0\pi^0$ cross section.                               
The target nucleus Coulomb field causes for the incoming                        
$\pi^+\pi^-$ pair enough shift of the {\em relative kinetic energy}             
towards higher energy values that                                               
the rapidly  falling transition cross section                                   
yields a reduced transition into the $\pi^0\pi^0$ channel and thus the          
surplus seen in experiment \cite{afa93}. Furthermore, we cannot confirm         
the used formulas in \cite{afa93}, especially the assumption of a small         
disturbance of the pure Coulomb atomic wave function by the                     
hadronic potential. Their                                                       
$A_{2\pi}$ lifetime is two orders of magnitude larger than                      
our calculations predict and the lifetimes increase according to our            
calculations with                                                               
$\tau_n=\tau_1 n^2$ and not  proportional                                       
$n^3$ as quoted in \cite{afa93}.                                                
\\                                                                              
Supported in part by Forschungszentrum J\"ulich, COSY Collaboration             
Grant No. 41126865.

                                                                                
\begin{table}\centering                                                         
\caption%
{$\pi\pi$ s--wave and $A_{2\pi}$ pionium ground  and                            
excited states.}                                                                
\label{tab_pipi}                                                                
\begin{tabular}[b]{cccccc}                                                      
 Model & $a_0^0 [\mu^{-1}] $ & $a_0^2 [\mu^{-1}] $ &                            
$E_B(n) [$keV$]$  & $\tau_n$ [s] & $\Gamma_n(FWHM)$ [eV]   \\                   
\hline                                                                          
 Froggatt                                                                       
& 0.31 &  $-0.059$ &       &                       &     \\                     
n=1 &  &           & 2.445 & $4.7 \times 10^{-17}$ & 14.0 \\                    
 Lohse                                                                          
& 0.30 &  $-0.025$ &       &                       &      \\                    
n=1 &  &           & 2.407 & $3.3 \times 10^{-17}$ & 20.3 \\                    
n=2 &  &           & 0.602 & $1.3 \times 10^{-16}$ & 5.06 \\                    
n=3 &  &           & 0.267 & $3.0 \times 10^{-16}$ & 2.25 \\                    
n=4 &  &           & 0.150 & $5.3 \times 10^{-16}$ & 1.27 \\                    
n=5 &  &           & 0.096 & $8.3 \times 10^{-16}$ & 0.81 \\                    
\end{tabular}                                                                   
\end{table}                                                                     
                                                                                
                                                                                
\unitlength1.0cm                                                                
\begin{figure}\centering                                                        
\begin{picture}(12.0,12.0)(0.0,0.0)                                             
\epsfig{figure=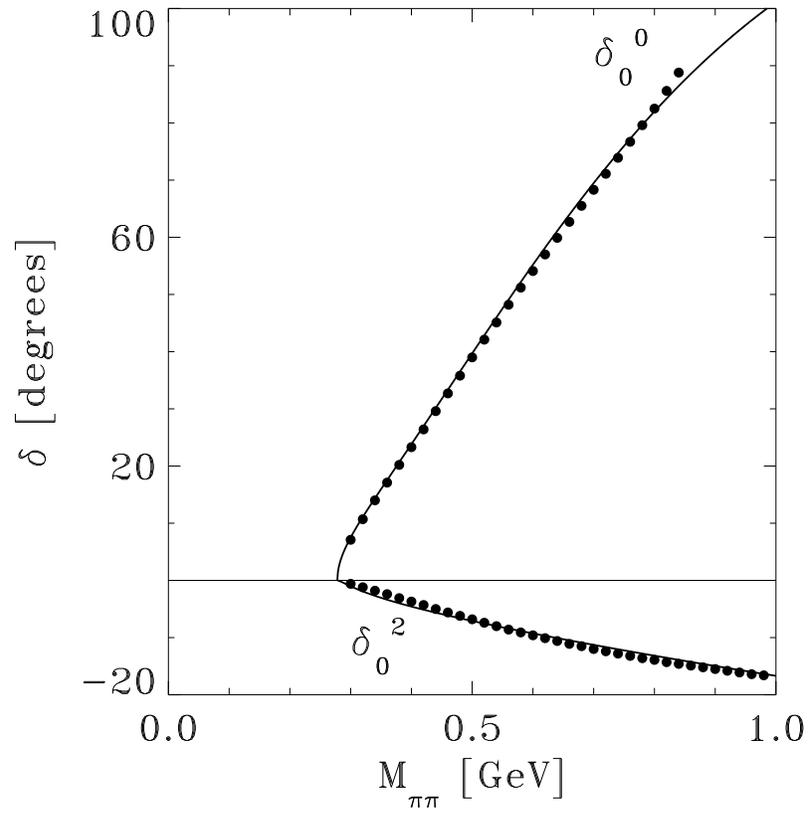,width=12.0cm}                                           
\end{picture}                                                                   
\caption%
{Froggatt data \protect\cite{fro77} (dots) with input                           
phase function (solid line).}                                                   
\label{fig1}                                                                    
\end{figure}                                                                    
                                                                                
\begin{figure}\centering                                                        
\begin{picture}(11.5,10.5)(0.0,0.0)                                             
\epsfig{figure=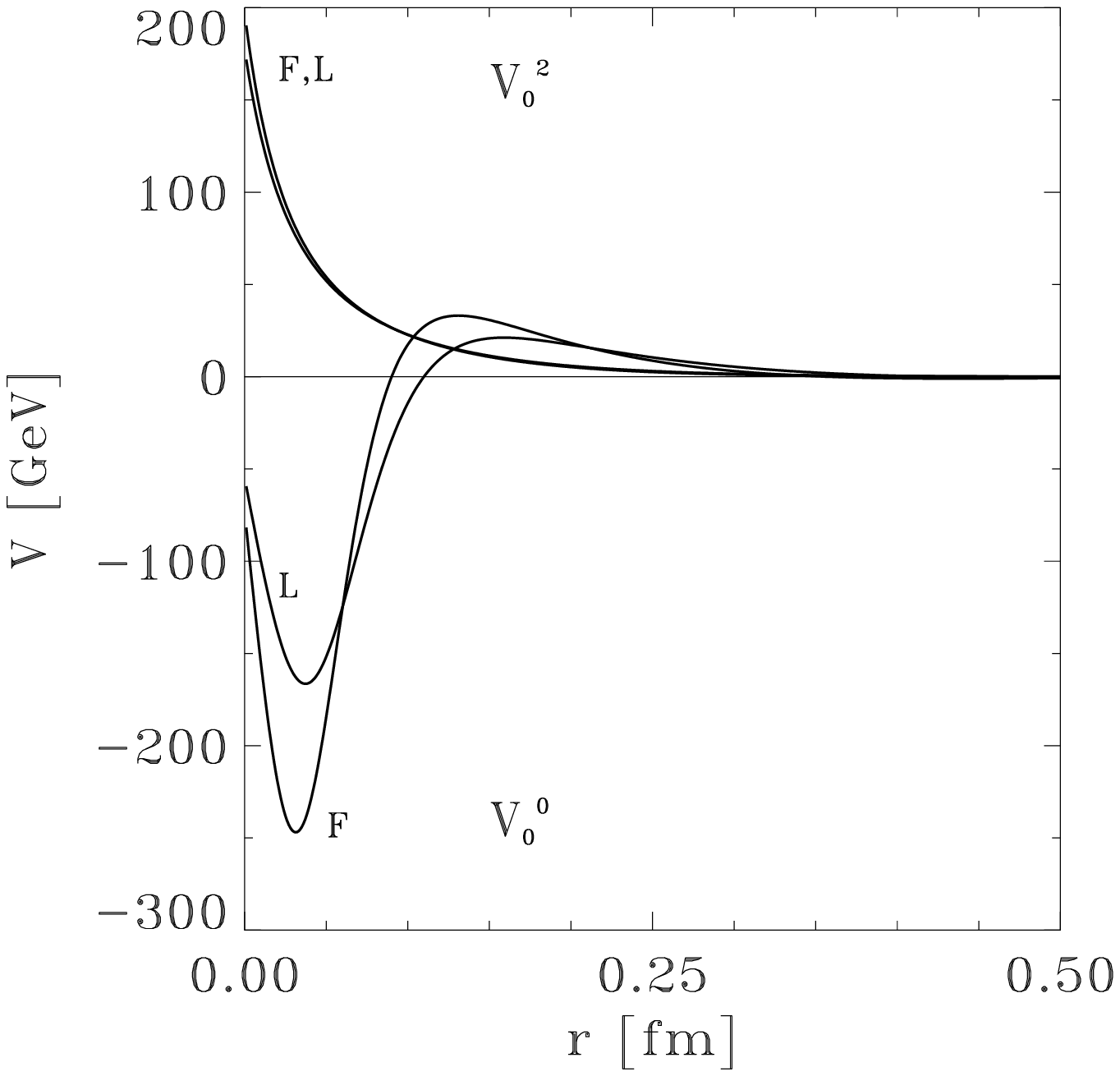,width=11.5cm}                                            
\end{picture}                                                                   
\begin{picture}(11.5,10.5)(0.0,0.0)                                             
\epsfig{figure=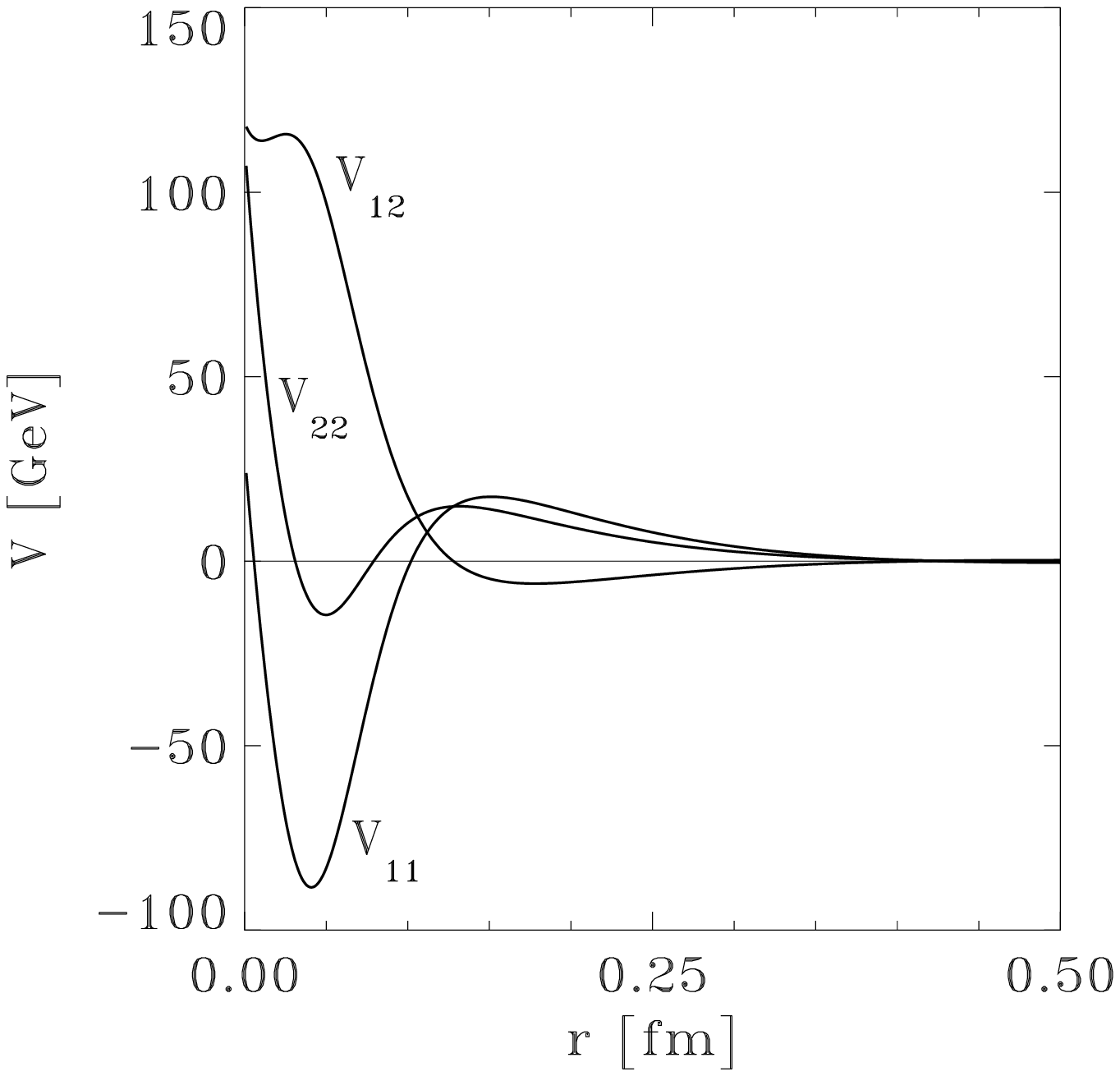,width=11.5cm}                                            
\end{picture}                                                                   
\caption%
{Inversion potentials based on Froggatt \protect\cite{fro77} and                
Lohse \protect\cite{loh90} phase shifts (top)                                  
and derived potential matrix                                                
from inversion of Lohse data \protect\cite{loh90} (bottom).}                    
\label{fig2}                                                                    
\end{figure}                                                                    
                                                                                
\begin{figure}\centering                                                        
\begin{picture}(11.5,10.5)(0.0,0.0)                                             
\epsfig{figure=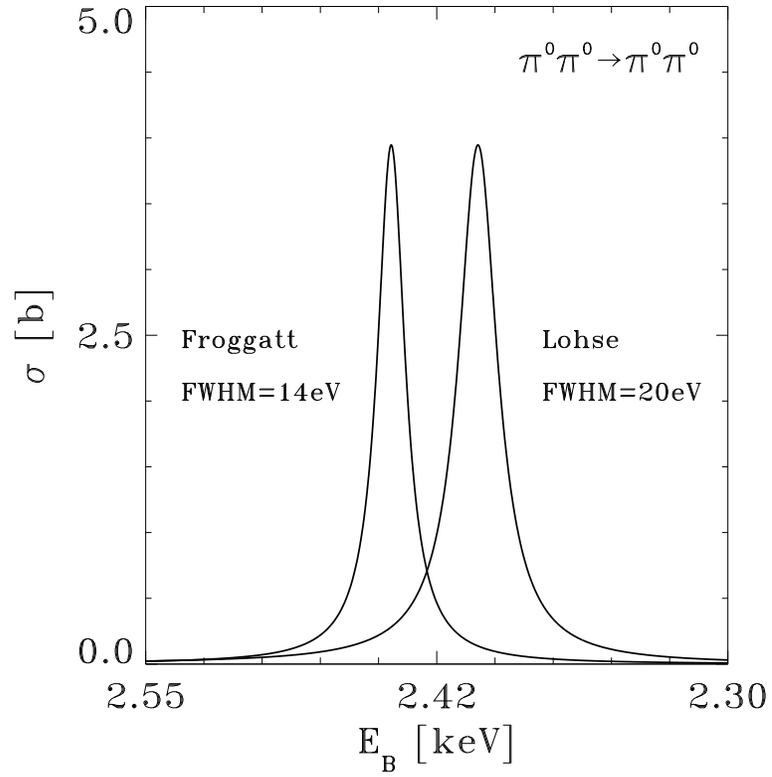,width=11.5cm}                                            
\end{picture}                                                                   
\begin{picture}(11.5,10.5)(0.0,0.0)                                             
\epsfig{figure=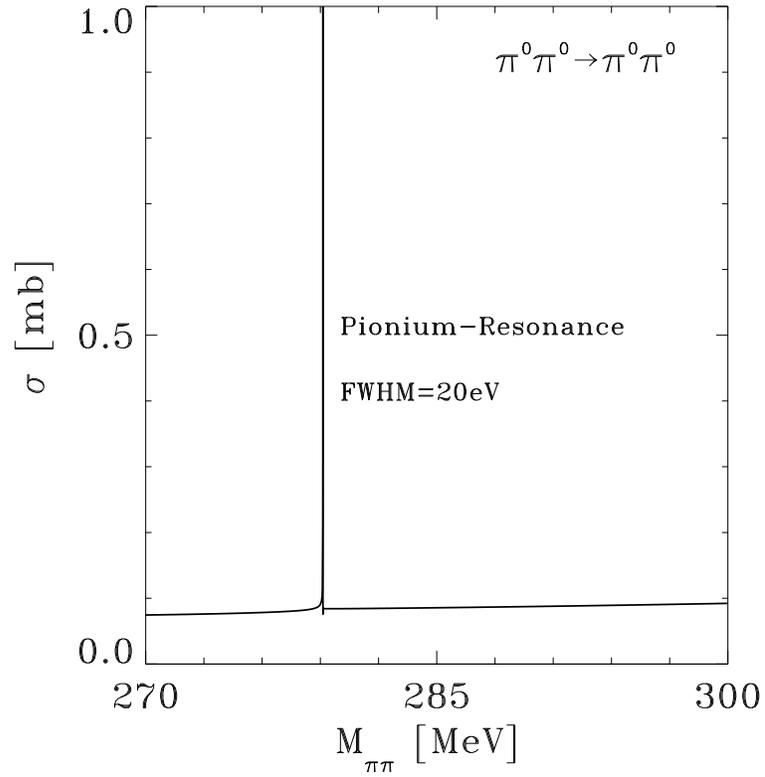,width=11.5cm}                                            
\end{picture}                                                                   
\caption%
{High resolution pionium proper resonance (top)
and general elastic cross section with $n=1$ resonance (bottom).}
\label{fig3}                                                                    
\end{figure}                                                                    
                                                                                
\begin{figure}\centering                                                        
\begin{picture}(12.0,12.0)(0.0,0.0)                                             
\epsfig{figure=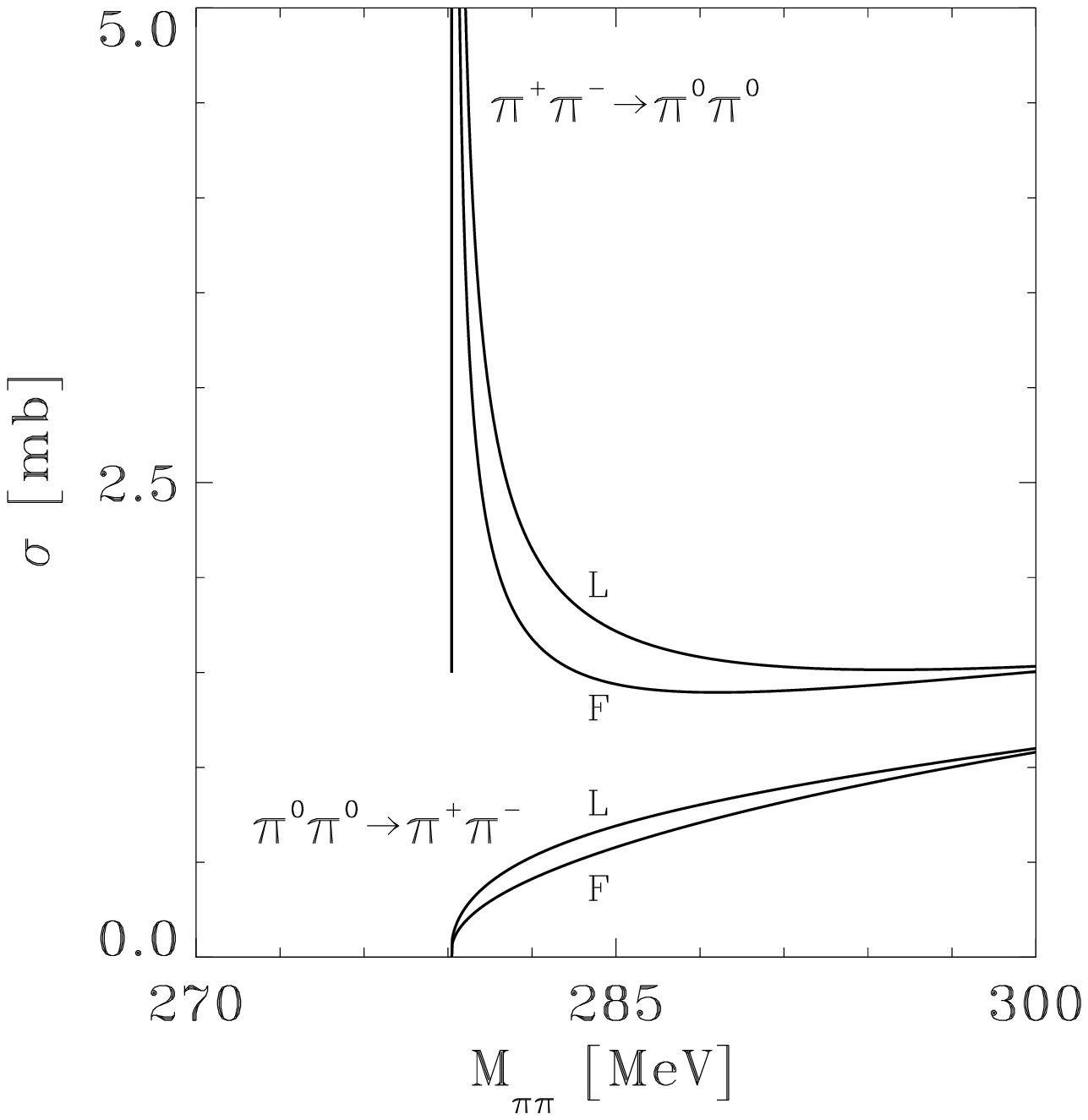,width=12.0cm}                                            
\end{picture}                                                                   
\caption%
{Transition cross sections based on Froggatt \protect\cite{fro77} and           
Lohse \protect\cite{loh90} phase shift inversion.}                              
\label{fig4}                                                                    
\end{figure}                                                                    

\begin{thebibliography}{99}                                                     
\bibitem{fro77}                                                                 
C.D. Froggatt and J.L. Petersen, Nucl. Phys. {\bf B129}, 89 (1977).             
\bibitem{loh90}                                                                 
  D. Lohse, J.W. Durso, K. Holinde, and J. Speth,                               
Nucl. Phys. {\bf A516}, 513 (1990).                                             
\bibitem{ger94}                                                                 
 H.V. von Geramb et al., in:                                                    
H.V. von Geramb (ed.), {\em Quantum Inversion Theory and Applications},         
 Lect. Notes in Physics 427, Springer (1994).                                   
\bibitem{afa93}                                                                 
 L.G. Afanasyev et al., Phys. Lett. {\bf B308},200 (1993),                      
{\em ibid.} {\bf B338}, 478 (1994).                                             
\bibitem{poc94}                                                                 
D. Po\v{c}ani\'{c}, in:                                                         
A.M. Bernstein and B.R. Holstein (eds.), {\em Chiral Dynamics: Theory and       
Experiment}, Lect. Notes in Physics 452, Springer (1995)                        
\bibitem{kur94}                                                                 
K. Kuroda et al., {\em Lifetime measurement of $\pi^+ \pi^-$ atoms},            
Proposal CERN--SPSLC--95--1, Dec. 1994                                          
                                                                                
\end{thebibliography}
\end{document}